# Performance Analysis of LMS Filter for SSPA Linearization in Different Modulation Conditions

J.N.Swaminathan[1], P.Kumar[2], M.Vinoth[2]

[1]*Chettinad College of Engineering & Technology,Puliyur-CF,Karur-63114,India*
[2]*K.S.Rangasamy College of Technology,Thiruchengode-637215,India*
[2]*Chettinad College of Engineering & Technology,Puliyur-CF,Karur-63114,India*

**Abstract**

The SSPA has wide application in Communication system, but its high output power varies due to its non linear gain. Pre-distortion method plays major role in power amplifier linearization. Polynomial is one of the methods used. The error estimation in Polynomial method is carried out by LMS Filter. Our main work is to analysis the error estimation performance of the LMS Filter for the Solid state power amplifiers (SSPA) in different modulation conditions. Here we are calculating the ACP and analyzing how effectively the memoryless non linearity has been reduced for all digital modulation techniques. All the analysis and results are taken using Matlab software.

## 1. Introduction

There are three types of linearization techniques to linearize the PA. 1. Feedback 2. Feed forward and 3. Pre-distortion [15] method. Feedback is not practically implemented since the error estimation process takes very long duration. Feed forward method is one of the existing methods, it achieves good linearization but it is having computational complexity due to complex algorithm in the RF end. Pre-distortion is also an existing method, It is simple and also having low computational complexity. It reduces the error to the acceptable level in the receiver end.

### 1.1. Pre-distortion - Polynomial method

The pre-distorter works like a magic box. It just consumes small power and reduces the ACP to a comparable level. The Pre-distortion concept can be implemented by 1. Look Up table method and 2. Polynomial method. In LUT method, it newly introduces the [5] quantization noise during the linearization process. The polynomial method measures the error using LMS filter algorithm and Pre-distorts [15] the error with input signal and gave the pre-distortion function which will be given as input to the SSPA. Due to the Pre-distortion function, an output with linear complex gain will be achieved.

The pre-distortion process will be explained as follows,
The output of the SSPA ($B_{out}$) after up conversion (modulation) giving the input signal $B_i$ is

$$B_{out} = a_1 B_i + a_2 B_i^2 + a_3 B_i^3 + a_4 B_i^4 + \ldots \ldots \ldots \qquad (1)$$

This is called as voltera series which explains the mathematical output of the PA. The output will be taken as a loop and down converted, here the even series signals (Harmonics) are filtered out and the resultant message signal is accompany with the inter-modulates.

$$B_{out}(t) = a_1 B(t + \tau_1) + a_3 B^3(t + \tau_3) + a_5 B^5(t + \tau_5) + \ldots \qquad (2)$$

With a delay of at the respective time period

*1.2. NLMS-Error estimation*

The inter-modulates are quantified by the slow adapt N-LMS algorithm filter. Here the step size of the filter plays the major role, If step size of the filter is high the inter-modulation estimation will be poor. So with low step size and slow adaptability conditions we construct the filter, which will estimates the error.

$$E(n) = - a_3 B^3(t + {}_3) - a_5 B^5(t + {}_5) + .... \quad (3)$$

From the estimated error the predistortion product is formed as $B_p$. the predistortion function then multiplied with the signal input $B_i$. The function of the predistorter behaves nearly opposite to the SSPA function. So it reduces all the non linearities and linearize the signal with linear complex gain.

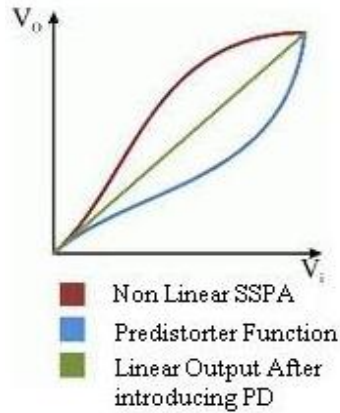

Fig.1. Non linear PA, Linear PA, Pre-distorter nature with respect to $B_i$ & $B_{out}$

Using the estimated error polynomial the pre-distortion function will be created which will be multiplied with the input $B_i$ gives the pre-distorted product. Here all of the pre-distortion functions are open looped and the $B_p$ product is given as input to the SSPA, so the resultant is a linearized output with complex gain.

## 2. SSPA Modeling

There are different types of power amplifier mathematical modeling for all memoryless non linear models. For the solid state power amplifier the ghorbani [1] modeling is highly suitable. The mathematical modeling will give the amplitude distortion [AM-AM] and phase distortion [AM-PM] mathematically which tell the mathematical behavior of the SSPA. The power amplifier modeling varies for different PA according to their gain function

$$H_{AM/AM} = \frac{x_1 h^{x_2}}{1 + x_3 h^{x_2}} + x_4 h \quad (4)$$

$$H_{AM/PM} = \frac{y_1 h^{y_2}}{1 + y_3 h^{y_2}} + y_4 h \quad (5)$$

$[x_1, x_2, x_3, x_4]$ are parameters used to compute the input amplitude gain [7] using equation (4). $[y_1, y_2, y_3, y_4]$ are the parameters used to compute the phase change using equation (5). The parameters inputs are given as [8.1081, 1.5413, 6.502, -0.0718] for gain and [4.6645, 2.0965, 10.88, -0.003] for phase change. The main application of solid state power amplifiers are in mobile communication and in radar communication which gives a output power of around 400watts with nonlinear gain.

*2.1. Digital Modulation techniques*

We performed pre-distortion process for four different digital modulation techniques [5][20] 1.16-QAM (Quadrature Amplitude Modulation) 2. 8-PSK (Phase Shift Keying) 3. QPSK (Quadrature Phase Shift Keying) 4. BPSK (Binary Phase Shift Keying). These modulation Techniques having different bandwidth. From lower bandwidth technique to higher bandwidth techniques are tabulated below in Table 1. As the spectral efficiency increases the complexity due to the non linearity also increases. We list out the Digital Modulations and its equations in a table along with their spectral efficiency.

Table 1. Different Digital Modulation Techniques

| Modulation | Modulation Equation | Bandwidth |
|---|---|---|
| 16-QAM | $X\sin(\omega_a t)\sin(\omega_c t)$ <br> $X = \pm 0.22$ | $F_b/4$ |
| 8-PSK | $X\sin(\omega_a t)\sin(\omega_c t)$ <br> $X = \pm 1.307$ | $F_b/3$ |
| QPSK | $\sin(\omega_a t)\sin(\omega_c t)$ | $F_b/2$ |
| BPSK | $\sin(\omega_a t)\sin(\omega_c t)$ | $F_b$ |

## 3. Linearization Performance

The Polynomial pre-distortion method for the SSPA linearization in different modulation conditions are executed successfully using Matlab. We came with quiet interesting results which are given below

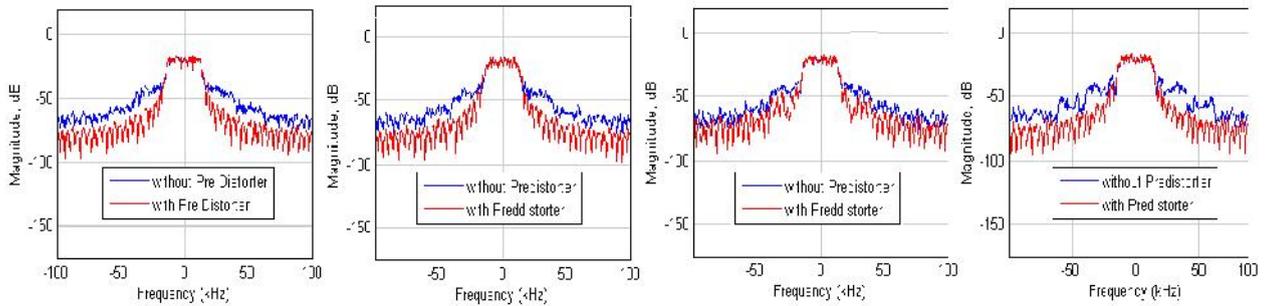

Fig. 2. Predistorter Performance for 16-QAM, QPSK, 8-PSK, BPSK

As earlier mentioned in section 1.2, due to the slow adapt conditions in the NLMS filter the linearization performance in for 16-QAM modulation is quiet good. Due to the predistorter all the inter-modulation products are get reduced to a considerable level. In QPSK & BPSK modulation technique ACP suppression is good for the IM3 only, the performance in these modulations can be improved by changing the step size of the filter, so a good noise suppression can be achieved .In 8-PSK a moderate level of improvement is achieved even in the slow adapt conditions. The ACP suppression level for each modulation is given below.

The above table shows, that there is much improved performance of the pre-distorter in 16-QAM, QPSK, BPSK. Due to the suppression of the inter-modulates, definitely the power level of the signal get improved for the Transmitter signal and also we can reduce the inter carrier interference due to the SSPA memoryless nonlinearity.

Table 2. Linearization Performance in Different Modulation Techniques

| Modulation | ACP Before Predistortion | ACP After Predistortion |
|------------|--------------------------|-------------------------|
| 16-QAM     | -28db                    | -52db                   |
| 8-PSK      | -29db                    | -35db                   |
| QPSK       | -29db                    | -47db                   |
| BPSK       | -28db                    | -49db                   |